\documentclass[journal,12pt,onecolumn,draftclsnofoot,]{IEEEtran}

\usepackage[table,xcdraw]{xcolor}
\usepackage{graphicx}
\graphicspath{ {img/} }
\usepackage[autosize]{dot2texi}
\usepackage{subcaption}
\let\labelindent\relax
\usepackage{enumitem}

\usepackage{tikz}
\usepackage[noadjust]{cite}
\usepackage{color}
\usepackage{multirow}
\usepackage[utf8]{inputenc}
\usepackage{pgfplots}
\usepackage{adjustbox}
\usepackage[utf8]{inputenc}
\usepackage{dirtytalk}

\begin{document}
\bstctlcite{IEEEexample:BSTcontrol}

\title{Mistral Supercomputer Job History Analysis}

\author{\small \IEEEauthorblockN{Micha{\l} Zasadzi{\'n}ski\IEEEauthorrefmark{1},
Victor Munt{\'e}s-Mulero\IEEEauthorrefmark{1}, Marc Sol{\'e}\IEEEauthorrefmark{1}and
Thomas Ludwig\IEEEauthorrefmark{3}}
\\
\IEEEauthorblockA{\IEEEauthorrefmark{1}CA Technologies, Barcelona, Spain, \{michal.zasadzinski, victor.muntes, marc.solesimo\}@ca.com\\
\IEEEauthorrefmark{3}Deutsches Klimarechenzentrum GmbH, Hamburg, Germany, ludwig@dkrz.de}}

\maketitle

\begin{abstract}
In this technical report, we show insights and results of operational data analysis from petascale supercomputer Mistral, which is ranked as 42nd most powerful in the world as of January 2018. Data sources include hardware monitoring data, job scheduler history, topology, and hardware information. We explore job state sequences, spatial distribution, and electric power patterns. \end{abstract}

\section{Introduction} \label{section_introduction}
The following report presents meaningful insights and results from data analysis of a huge computing environment - supercomputer Mistral\footnote{https://www.dkrz.de/up/systems/mistral}, ranked as 42nd most powerful on the world as of January 2018\footnote{Ranking November 2017 https://www.top500.org/system/178567}. The HPC system has a peak performance of 3.14 PetaFLOPS and consists of approx. 3,300 compute nodes, 100,000 compute cores, 266 Terabytes of memory, and 54 PiB of Lustre file system. Presented work is associated with  PhD research on root cause analysis for complex and distributed IT systems \cite{Zasadzinski:2017, Zasadzinski:2016fast}.

\section{Data center of German Climate Computing Centre (DKRZ)} \label{section_mistral}
Data center placed in DKRZ contains (1) Mistral supercomputer, which comprises 3336 computing nodes placed in 47 racks, (2) about 90 special nodes dedicated for maintenance activities, data pre-processing, post-processing and advanced visualizations, (3) a separate 54 petabytes Lustre file system. Majority of the racks are homogeneous having mounted the same 72 blades, and each rack encloses 4 or 2 chassis, with the maximum capacity of 18 blades per chassis. Computing nodes are divided into several partitions: development, pre/postprocessing, test, production. In this paper, we analyze data from the production partitions. The workload is generated by a variety of applications and simulators used in areas such as climate science, geology, and natural environment. In this data center, resource allocation and accounting are maintained using Slurm\footnote{https://slurm.schedmd.com/}. This open source Resource and Job Management System manages nodes' reservations from users. The computing nodes are Intel Xeon 12C 2.5GHz (Haswell) and Intel Xeon 18C 2.1GHz (Broadwell). 

Chassis contains 18 blades, less frequently 12 or 16. Understanding the data center structure is essential for the creation of diagnostic models. First of all, it allows determining the structures from which we need to collect data. Secondly, it enables the detection of local interactions between neighboring devices. For example, in the Mistral system, inventory tables contain detailed information about all the installed equipment, its interconnections, management controllers, and localization. Table \ref{table:computing_nodes} presents the number of computing nodes by type, e.g., B720\_24\_64 stands for Bull B720 with 24 cores and 64 GB of RAM. 
Furthermore, the used dataset contains metrics and topology information from other equipment installed in the data center, such as controllers, switches, power meters, water valves, cold doors. See Table \ref{table:computing_nodes} for details. The rest of the racks contain blades of diverse types, but in each chassis, there is exactly one node type.

\begin{table}[h]
\centering
\caption{Computing nodes in Mistral}
\label{table:computing_nodes}

\begin{tabular}{|l|r|r|ll}
\cline{1-3}
\textbf{Node type}& \textbf{Quantity} & \textbf{Homogeneous racks, 72 blades}\\ \cline{1-3}
 B720\_36\_64  & 1454     & 20\\ \cline{1-3}
 B720\_24\_64  & 1404     & 19\\ \cline{1-3}
 B720\_36\_128 & 270      & 3\\ \cline{1-3}
 B720\_24\_128 & 110      & 1\\ \cline{1-3}
 B720\_36\_256 & 50       & -\\ \cline{1-3}
 B720\_24\_256 & 48       & -\\ \cline{1-3}
\end{tabular}
\end{table}

\subsection{Job scheduler history} \label{subsection_job_scheduler_history}
According to the dataset, each job submission finishes with one of the following states, defined by Slurm documentation.
\begin{itemize}
 \item \textbf{Completed} -- Job has terminated all processes on all nodes with an exit code of zero.
 \item \textbf{Cancelled} -- Job is canceled by the user or administrator. The job may or may not have been initiated. In the following analysis, we take into account only cancelled jobs longer than 0 s
  \item \textbf{Timeout} -- Job terminated upon reaching its time limit.
 \item \textbf{Failed} -- Job terminated with non-zero exit code or other failure condition. According to Mistral, other failure condition includes failures caused by any external factor to an allocated node, e.g., failures of Lustre FS, IB.
 \item \textbf{Node fail} -- Job terminated due to a failure of one or more allocated nodes. This state includes only own hardware related problems of a computational node.
\end{itemize}

Users submit jobs. These jobs consist of one or more steps. Steps are sets of tasks within a job. Steps execution order is defined, and they may be executed sequentially, in parallel, or mixed. However, most steps in Mistral dataset are executed sequentially. For instance, a single step may utilize all nodes allocated to the job, or several steps may independently use a portion of the allocation. In the Slurm database, there are 76 columns with information about each job. Part of these columns includes the job configuration specified by a user, such as allocated nodes, time limit, required CPU frequency. Others contain job statistics, which include timing, average hardware usage, such as disk read/write (R/W) -- the sum of local storage and Lustre operations done by a particular job such as virtual memory (VM) size. In this paper, we consider all listed job states. \textbf{For steps, the dataset only includes: Completed, Failed and Cancelled.}

\begin{figure}[!h]
\centering
\includegraphics[width=0.5\textwidth]{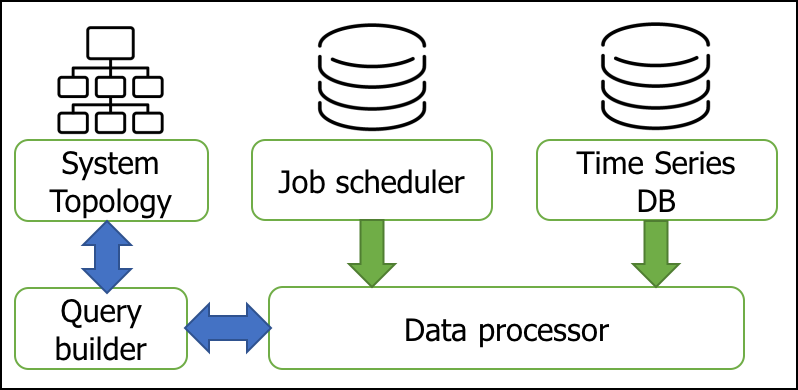}
\caption{ High-level data processing scheme}
\label{schema_v2}
\end{figure}

In Figure \ref{schema_v2}, a high-level scheme of data processing modules and data sources is presented.
\vspace{-10pt}
\section{Failed Job Analysis} \label{section_case_study}

\subsection{General statistics} \label{general_stats} According to the data from the job scheduler, more than 1.3M jobs, and more than 270k different job names were submitted in the 10-months period represented by the dataset extracted from the Mistral production environment. These submissions, which are mainly executed in batch mode (98.8\%), resulting in over 4.8M steps. Completed jobs are 91.3\% of all submitted ones. In contrast, 5.6\% of started jobs result in the fail state, 1.7\% of submissions are canceled, 1.4\% result in timeout, and 0.028\% fail because of a computing node problems. Through the analysis of these data, it is observed that the mean number of allocated nodes is 3.4 for completed steps and 18 for failed ones. This result follows a typical pattern usually reported in state of the art: failed steps are usually more complicated. Average duration and standard deviation of failed jobs and completed ones are quite similar. When it comes to steps, completed ones take in average \(414 s\), while failed almost three times more.
For detailed statistics, see Table \ref{main_jobs} for jobs and Table \ref{table_job_steps} for steps. These general statistics represent a convincing motivation for generating savings with the early termination of jobs predicted to fail. An average failed job consumes many more CPU hours than completed one and decreases resources availability. About 1.2M of all steps from the set run for more than \(60 s\) and 1.1M more than \(120 s\).


\setlength{\tabcolsep}{3pt}

\begin{table}[h]
\centering
\caption{Jobs statistics by Slurm state}
\label{main_jobs}
\begin{adjustbox}{width=0.8\textwidth}

\begin{tabular}{l|l|l|l|l|l|l|l|l|l|l|l|l|l}
\hline
\multirow{2}{*}{State} & \multirow{2}{*}{count} & \multicolumn{6}{l|}{Allocated nodes}     & \multicolumn{6}{l}{Duration {[}s{]}}         \\ \cline{3-14} 
                       &                        & mean & SD & min & 50\% & 75\% & max  & mean  & SD & min & 50\% & 75\%  & max     \\ \hline \hline
CANCELLED              & 23087                  & 25   & 100    & 1   & 5    & 16   & 3264 & 2680  & 14952  & 1   & 310  & 1591  & 1.6M \\ \hline
COMPLETED              & 1238585                & 12   & 34     & 1   & 6    & 16   & 3276 & 1954  & 4190   & 1   & 419  & 1953  & 0.3M  \\ \hline
FAILED                 & 75897                  & 15   & 39     & 1   & 6    & 16   & 1700 & 1763  & 4288   & 1   & 164  & 2979  & 0.4M  \\ \hline
NODE\_FAIL             & 390                    & 67   & 289    & 1   & 10   & 46   & 3264 & 11087 & 105332 & 38  & 2472 & 6202  & 2.1M \\ \hline
TIMEOUT                & 17864                  & 16   & 57     & 1   & 1    & 16   & 3078 & 11586 & 18140  & 60  & 2408 & 28803 & 0.6M  \\ \hline
ALL                    & 1355823                & 13   & 37     & 1   & 6    & 16   & 3276 & 2085  & 5444   & 1   & 425  & 2001  & 2.1M \\ \hline
\end{tabular}
\end{adjustbox}
\end{table}

\setlength{\tabcolsep}{1pt}

\setlength{\tabcolsep}{3pt}
\begin{table}[h]
\centering
\caption{Steps statistics by Slurm state}
\label{table_job_steps}
\begin{adjustbox}{width=0.9\textwidth}

\begin{tabular}{l|l|l|l|l|l|l|l|l|l|l|l|l|l|l|l|l|l}
\hline
\multirow{2}{*}{State} & \multirow{2}{*}{count} & \multicolumn{4}{l|}{\textbf{Allocated nodes}} & \multicolumn{4}{l|}{\textbf{Duration {[}s{]}}} & \multicolumn{4}{l|}{\textbf{Ave Disk Read {[}GB{]}}} & \multicolumn{4}{l}{\textbf{Ave Disk Write {[}GB{]}}} \\ \cline{3-18} 
                       &                        & mean      & SD      & min      & max      & mean      & SD      & min      & max       & mean        & SD        & min       & max        & mean        & SD        & min       & max         \\ \hline \hline
CANCELLED              & 53579                  & 28        & 87          & 1        & 3264     & 3322      & 8679        & 1        & 183k      & 16          & 151           & 0         & 7821       & 3           & 37            & 0         & 2341        \\ \hline
COMPLETED              & 4853842                & 3.4       & 17          & 1        & 3276     & 414       & 1902        & 1        & 235k      & 1           & 10            & 0         & 6993       & 0.2         & 5             & 0         & 30742       \\ \hline
FAILED                 & 197704                 & 18        & 28          & 1        & 3249     & 1111      & 5273        & 1        & 346k      & 3           & 73            & 0         & 6629       & 0.2         & 15            & 0         & 4078        \\ \hline
ALL                    & 5105125                & 4.2       & 20          & 1        & 3276     & 471       & 2326        & 1        & 346k      & 1           & 23            & 0         & 7821       & 0.2         & 7             & 0         & 4078        \\ \hline
\end{tabular}
\end{adjustbox}
\end{table}

\subsection{Job state sequences}
\begin{figure}[!h]
\centering
\includegraphics[width=0.7\textwidth]{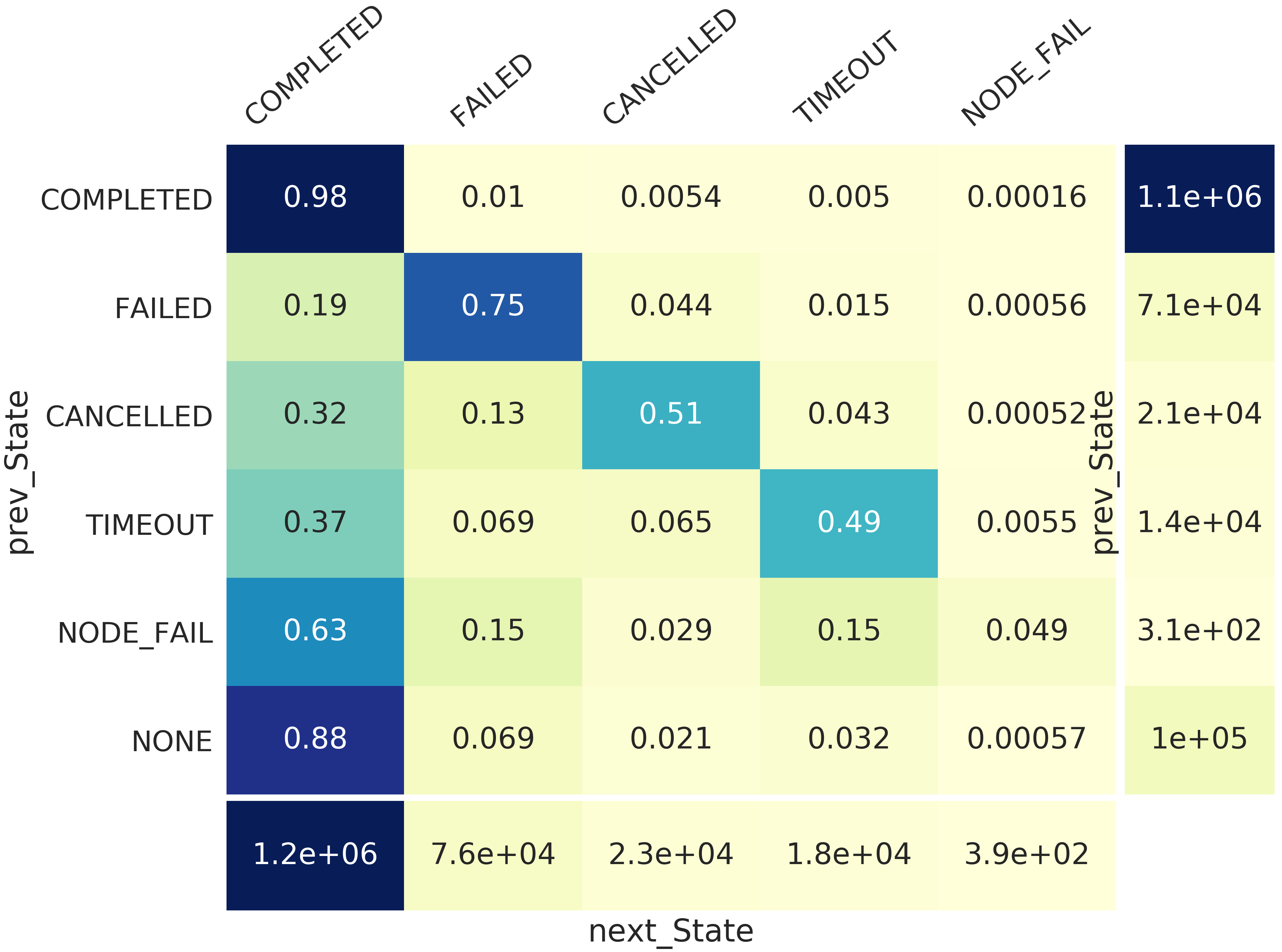}
\caption{Heat-map presenting transition between 2 subsequent jobs, grouped by a user name and job name. For instance, after 0.32 of all jobs which are canceled, the next jobs are completed ones}
\label{heatmap_state_transition}
\end{figure}

Outcomes from previous analysis encourage the analysis of correlations between user's past jobs and the final state of a subsequent job. Firstly, we create a matrix presenting job state transitions. In details, Figure \ref{heatmap_state_transition} illustrates states of 2-jobs sequences, grouped by a user name and job name (exact string match). 
Another possibility to build these sequences is to match jobs by parts of their names, e.g., without suffixes, which usually stand for a simulated year, or another parameter of a run application. 
Previous state NONE refers to initial submissions, from which 88\% completes, and the majority of the rest fails. Importantly, only 19\% of next submissions complete after a job failed and 75\% of them still fail. Majority of jobs completes after a hardware failure of a node. Also, these data reveal a few interesting rationales. For instance, users often submit applications which are correct and do not fail. Then they start trails, implement changes, or just merely develop their models. Majority of next submissions completes, but still, failures are two times more probable than cancellations or timeouts. A typical user is more likely to have a job in the completed state after it is canceled than it is failed. An interesting fact is that the probability of a node failure reaches its maximum value after another node failure, and it has the same order of magnitude for all other states.

\begin{figure}[!h]
\centering
\includegraphics[width=0.8\textwidth]{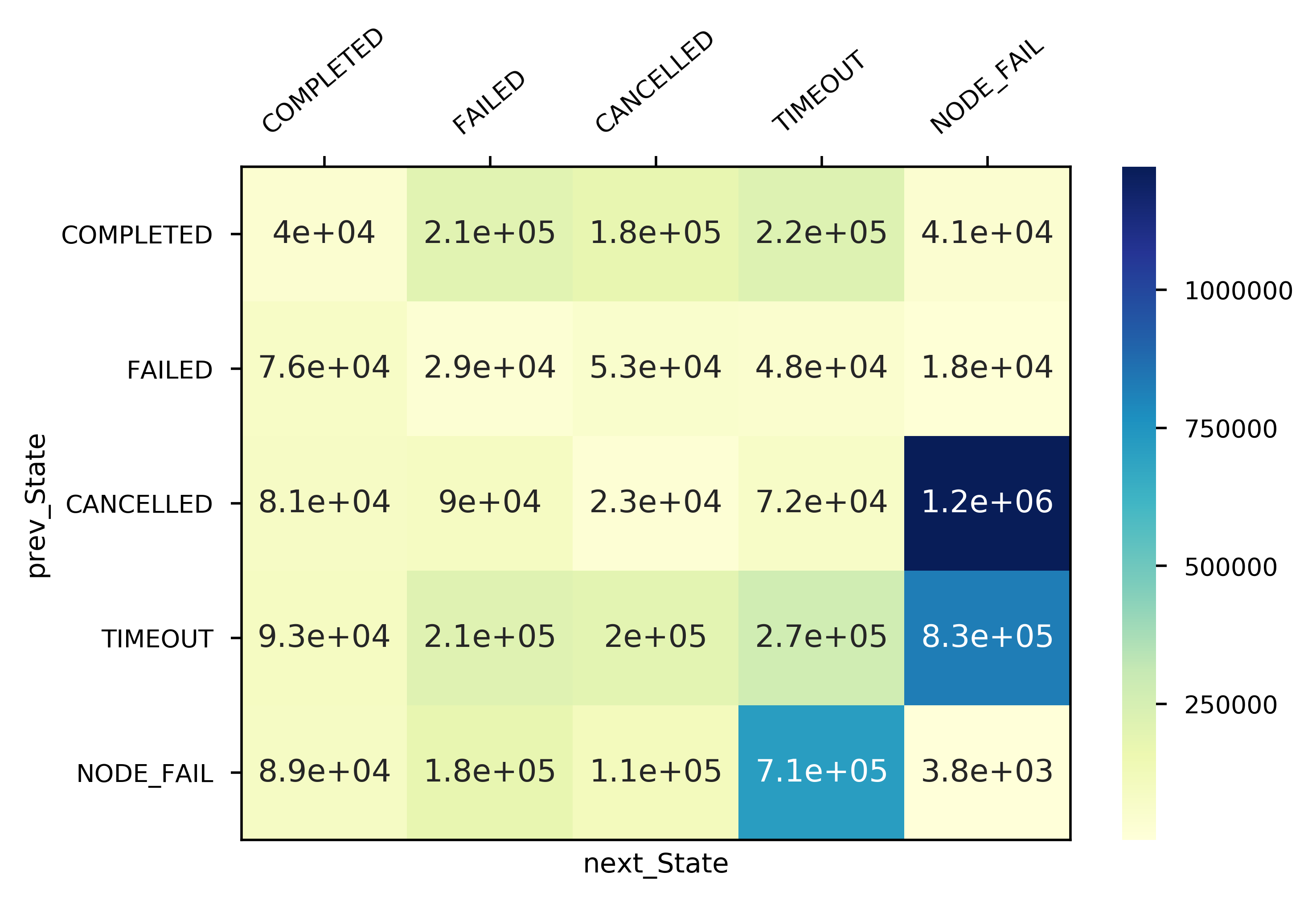}
\caption{Heat-map presenting mean time [in seconds] between subsequent job states, grouped by user, application name.}
\label{heatmap_state_transition_time}
\end{figure}

\begin{figure}[!h]
\centering
\includegraphics[width=0.8\textwidth]{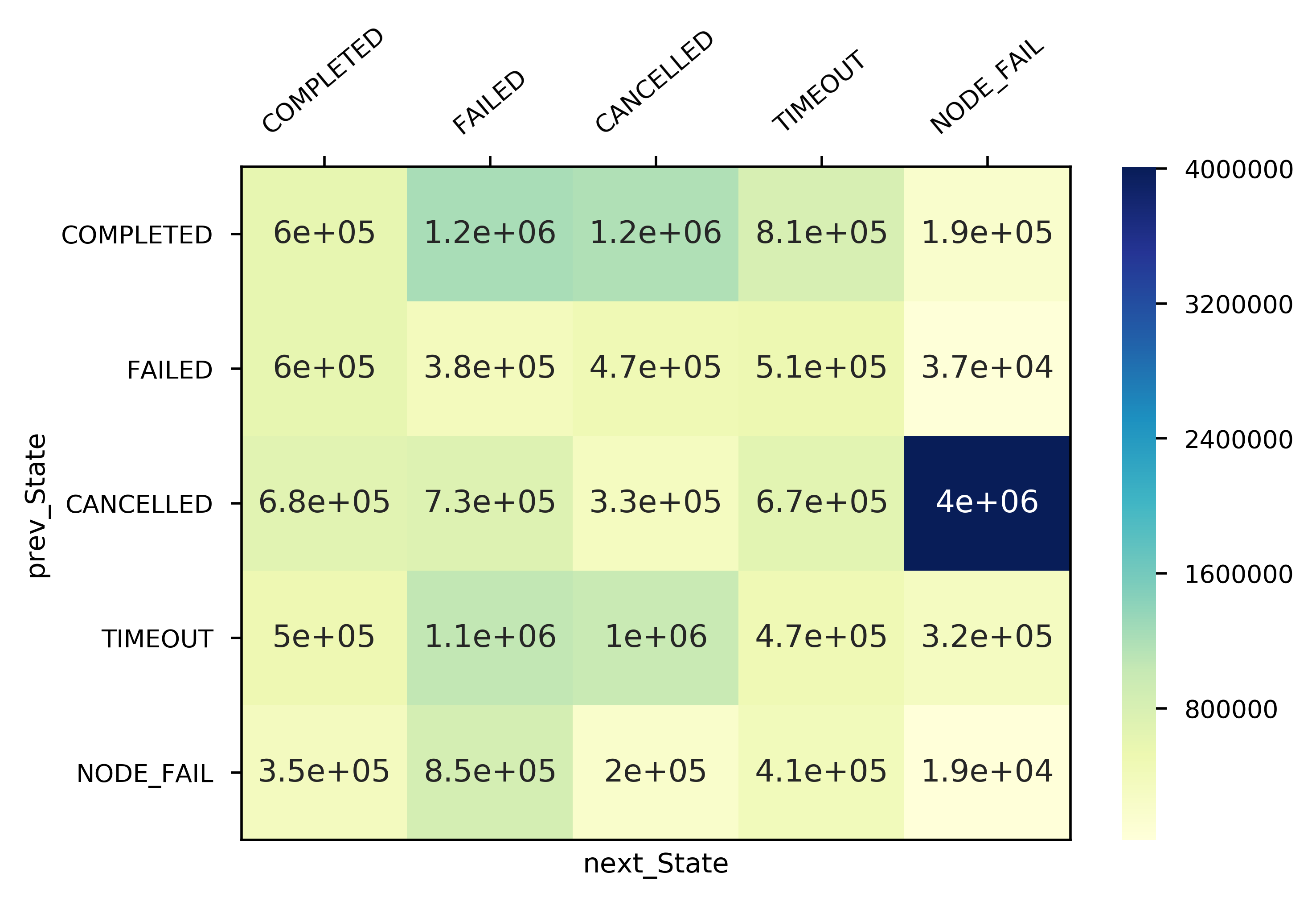}
\caption{Heat-map presenting SD [in seconds] between subsequent job states, grouped by user, application name.}
\label{heatmap_state_transition_time_stddev}
\end{figure}
\begin{figure}[!h]
\centering
\includegraphics[width=0.6\textwidth]{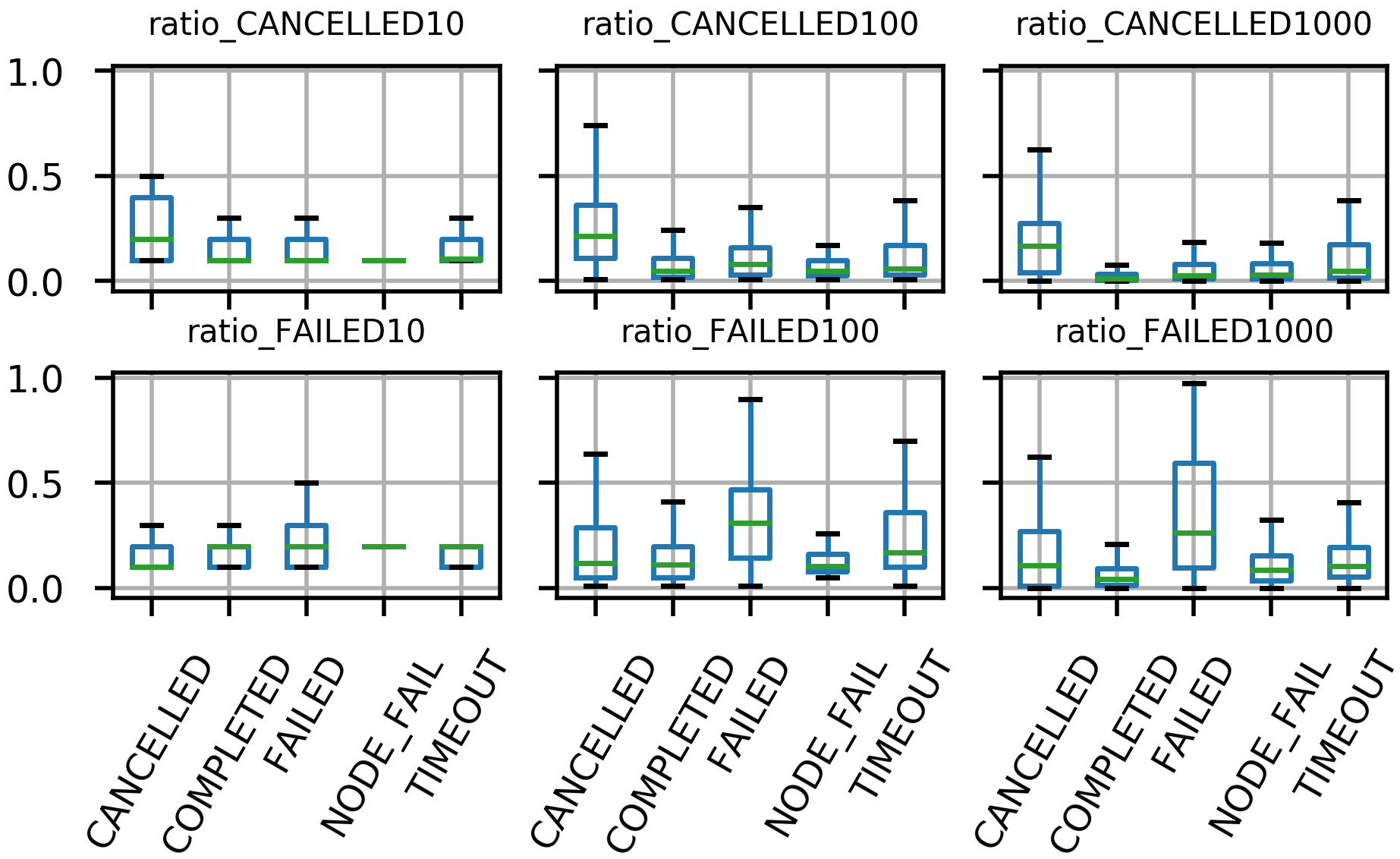}
\caption{Plot presenting distributions of users' factors of failed and cancelled jobs for last N=10, 100, 1000 submissions for each succeeding job state. Users with more than 10 jobs submitted are counted. For instance, before failed job, a max. factor of failed in window of last N=10 submissions is 0.5.}
\label{last_canc_fail_vs_state}
\vspace{-20pt}

\end{figure}

Regarding the correlation between cancelled and failed, 13\% of next submissions after cancellations fails and only one third completes. Moreover, Table \ref{table_job_steps} shows that cancelled steps are characterized by much higher disk RW than completed and even failed ones. One of the potential causes after interviewing system administrators is that they cancel steps, due to high storage system usage -- IO counters. Obviously, after cancellation, a job is possibly corrected and re-submitted to be completed. Further analysis is shown in Figure \ref{last_canc_fail_vs_state} which presents average factors of past failed and cancelled jobs to all submitted jobs in different N number of prior submissions for each job state. A readable observation is that, on average, in preceding ten jobs there are as many cancelled jobs as failed ones for all states except node fail - probably lack of diverse samples. It can be highlighted that a cancellation often follows up other cancellations and a failure other failures.

\begin{figure}[!h]
\centering
\includegraphics[width=0.7\textwidth]{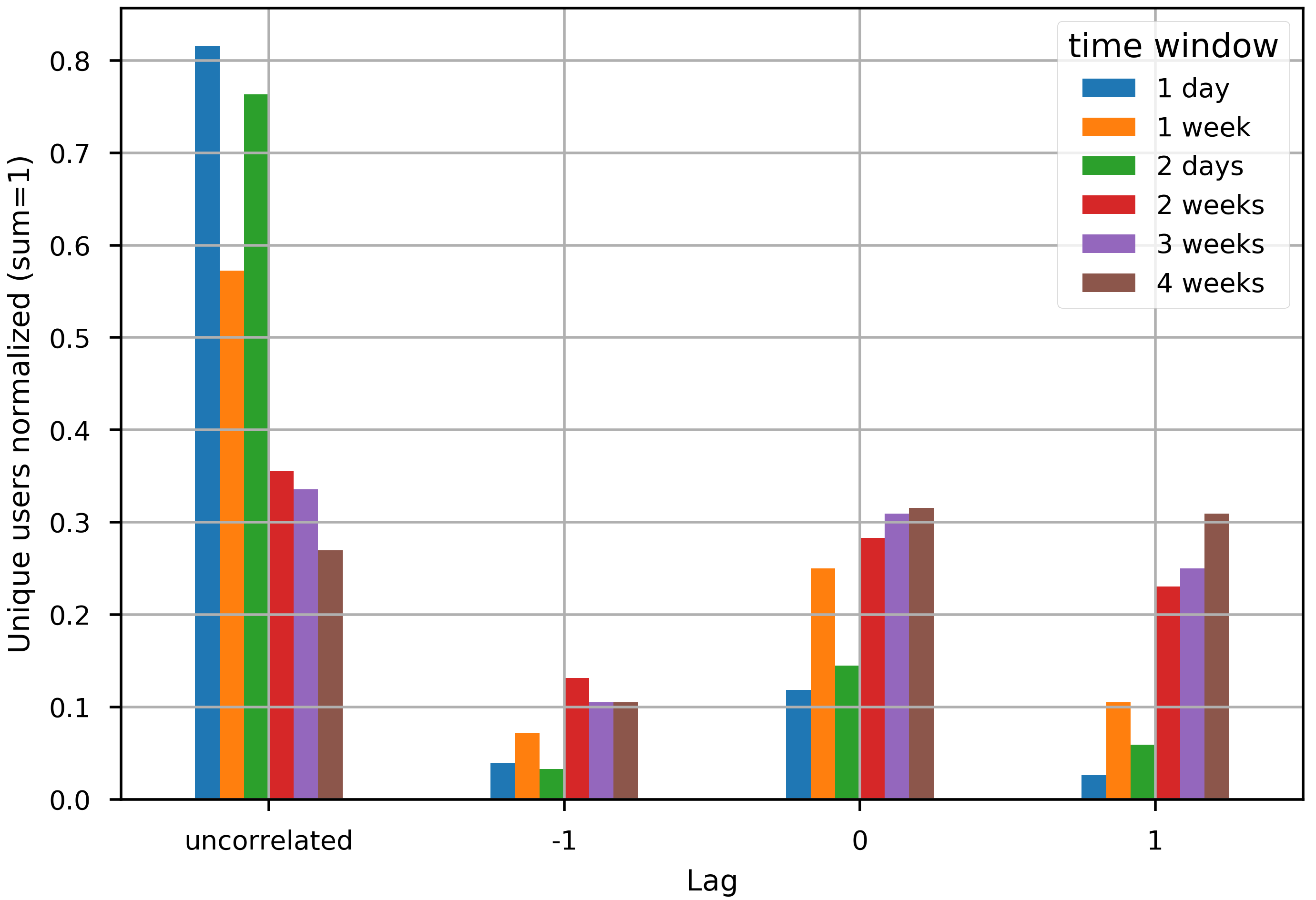}
\caption{Plot presenting distribution of Pearson correlation coefficient for users with min. 1000 jobs submitted, correlation is counted for coefficients \(> 0.3\). Total 304 users.}
\label{corr_vs_lags}

\end{figure}

\begin{figure}[!h]
\centering
\includegraphics[width=0.7\textwidth]{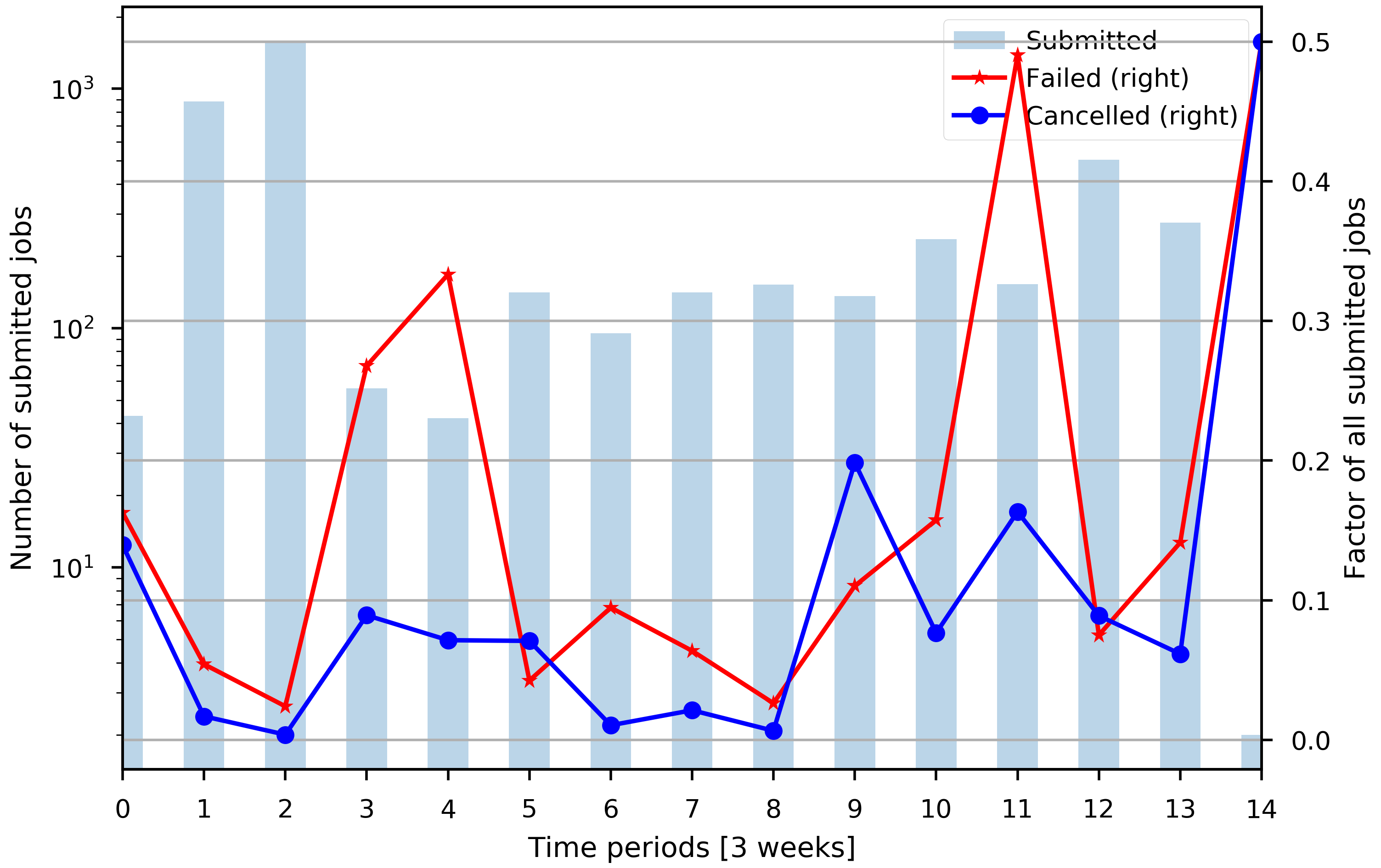}
\caption{Plot presenting cancelled and failed job sequences aggregated in 4-week periods, for a relatively active user.}
\label{cancelled_failed_sum}

\end{figure}

\begin{figure}[!h]
\centering
\includegraphics[width=0.7\textwidth]{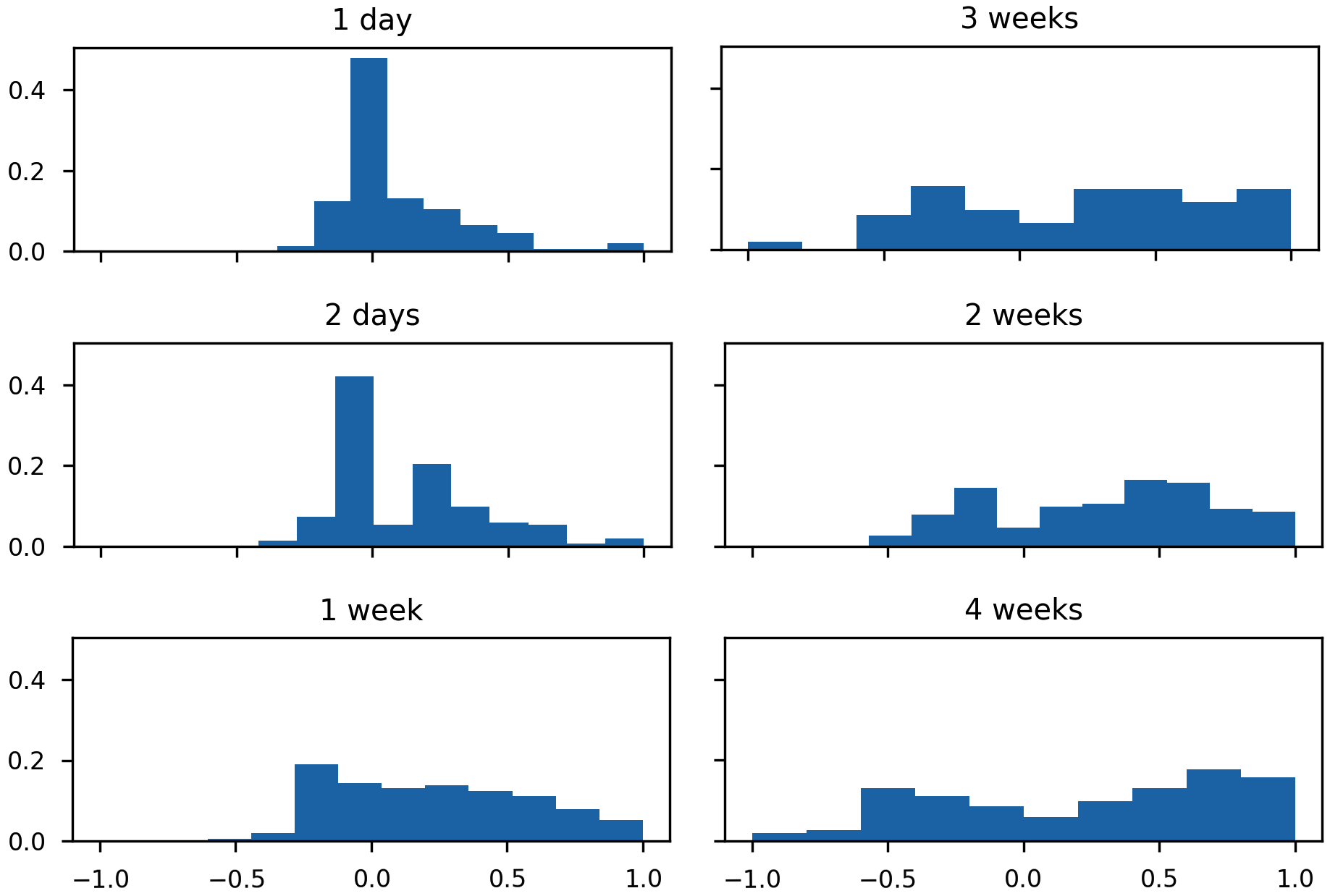}
\caption{Plot presenting cancelled and failed jobs Pearson correlation coeff. distribution for users by aggregation periods}
\label{hist_corrs}
\vspace{-20pt}

\end{figure}

Besides, in Figure \ref{corr_vs_lags}, we present correlation type distribution between a number of failed and cancelled jobs in different time windows. Aggregation in 4-week periods and no lag between these sequences reveals the highest number of sequences with correlation coefficient over a fixed threshold of 0.3. Additionally, we present distributions of a correlation coefficient value, see Figure \ref{hist_corrs} for different time windows. These distributions show that correlations are stronger for longer periods -- weeks over days.
In link with this, sequences of cancellations and failures are presented in Figure \ref{cancelled_failed_sum} for a randomly chosen user with relatively high activity. Surprisingly, it is observed that local minima of failed and cancelled jobs exist in the same time periods. In contrast, high activity of a user does not necessarily mean a high number of failures and cancellations. Naturally, a user might submit the same working code. These sequences reveal that there are periods of re-running the same models, and periods of experiments when a model is changed. This phenomenon is confirmed by researchers working in DKRZ.

\subsection{Time view}
\begin{figure}[!h]
\centering
\includegraphics[width=0.9\textwidth]{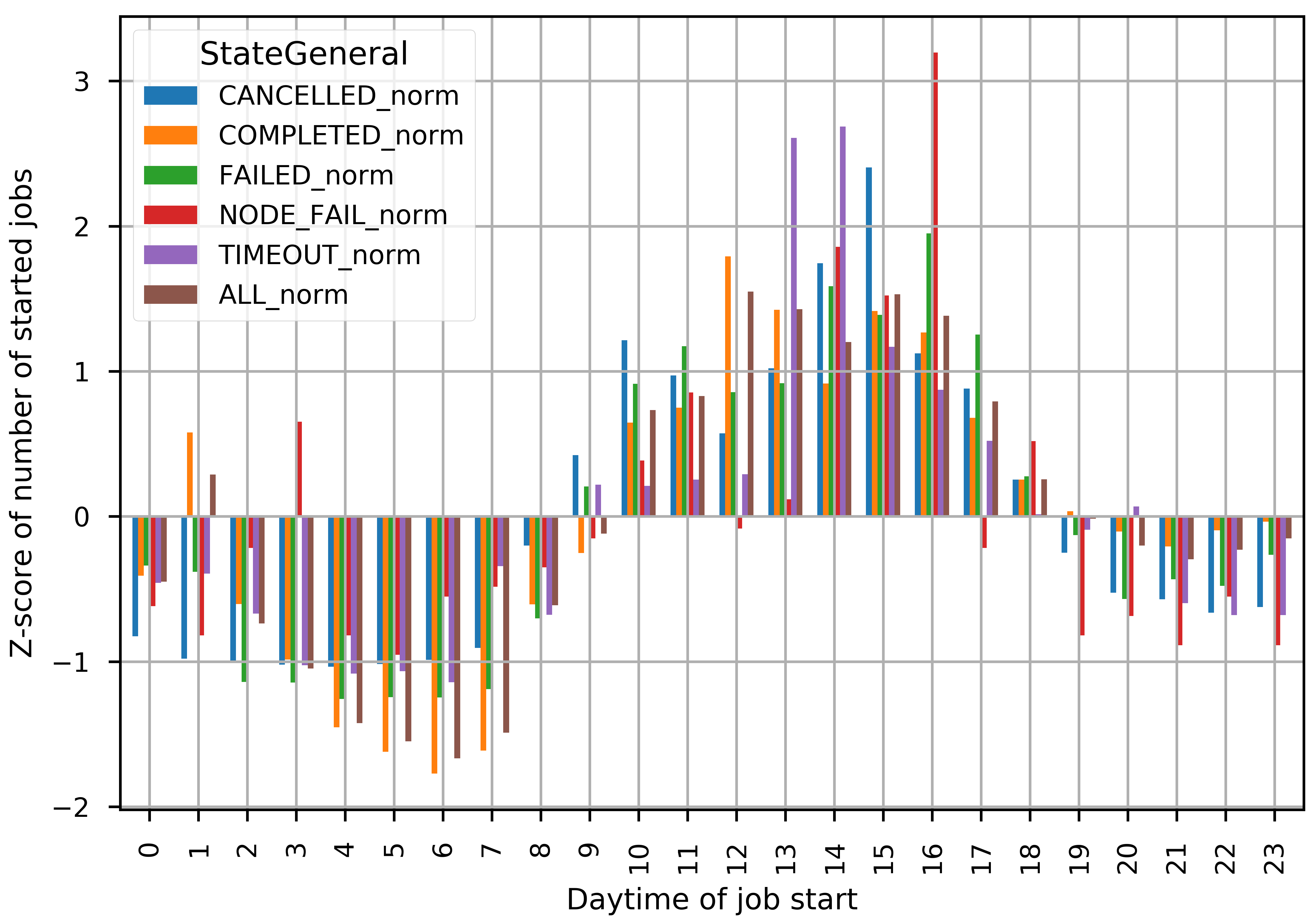}
\caption{Plot presenting cancelled and failed jobs depending on the daytime of start}
\label{start_daytime_norm}
\vspace{-30pt}

\end{figure}

\begin{figure}[!h]
\centering
\includegraphics[width=0.7\textwidth]{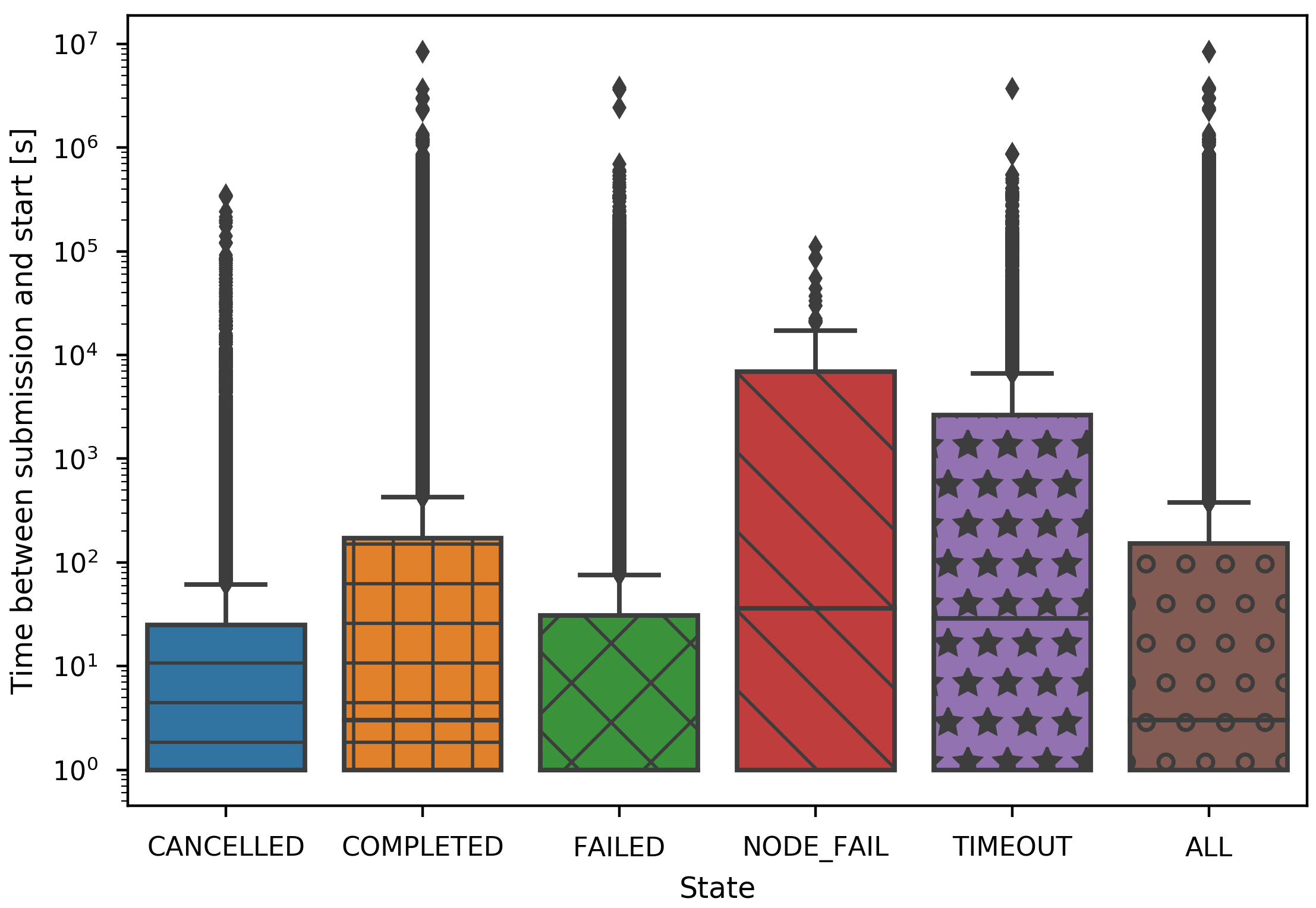}
\caption{Plot presenting distributions of waiting time of submissions by state}
\label{plot_waiting_time}
\vspace{-30pt}
\end{figure}

The overall cycle of jobs depending on the daytime can be seen in Figure \ref{start_daytime_norm}. The number of jobs by the state is normalized to the mean number of started jobs during the whole daytime. Naturally, during the night the number of started jobs is much lower. Between 10 and 17 hours, the number of submissions is over the mean. Moreover, in Figure \ref{plot_waiting_time}, we present distribution of time elapsed from job submission to a job start. This distribution shows that the highest waiting time is for jobs resulted in a timeout and node fail state.

\begin{figure}[!h]\centering\includegraphics[width=0.7\textwidth]{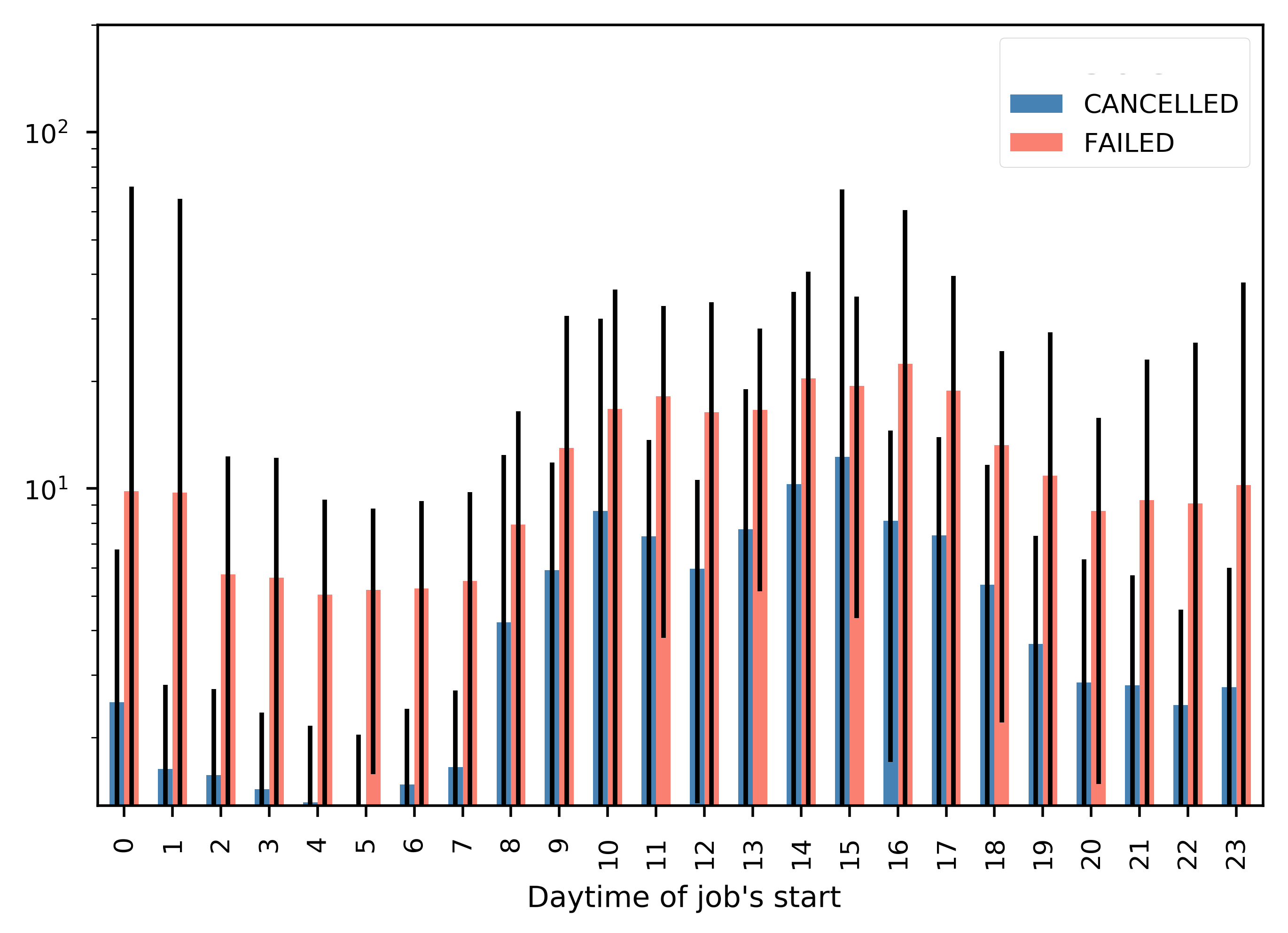}\caption{Plot presenting daily mean number (with stddev) of jobs finished as cancelled or failed by the daytime of start}\label{cancelled_failed_daytime}

\end{figure}

In Figure \ref{cancelled_failed_daytime}, we present the average number of cancelled and failed jobs aggregated by daytime. It is clear that the highest number of failed ones starts between 14 and 16, while for cancelled the maximum is at 15 hour.

\subsection{Distribution of a job over the data center}
The job scheduler is optimized to use nodes which are closest to each other to reduce latency in data transfer. Topology-aware resources allocation is applied as well as in Slurm, and other schedulers.
An interesting aspect to explore might be the distribution of the jobs over racks. 
Through this, we can discover the dependency between the number of network hops and failed jobs. The number of hops represents the complexity of a network topology for a particular job and increases with the number of used racks since a switch is mounted in each chassis. For this, we choose a subset of steps allocated on more than one node with duration more than \(60 s\). In average, completed steps are allocated on 1.1, \(\sigma=0.8\) racks, cancelled on 2.3, \(\sigma=2.8\) and failed on 1.8, \(\sigma=1.8\). Completed steps are not only distinguished by the lowest number of used racks, but also the lowest number of allocated nodes, as seen in Table \ref{table_job_steps}.

\begin{figure}[!h]
\centering
\includegraphics[width=0.7\textwidth]{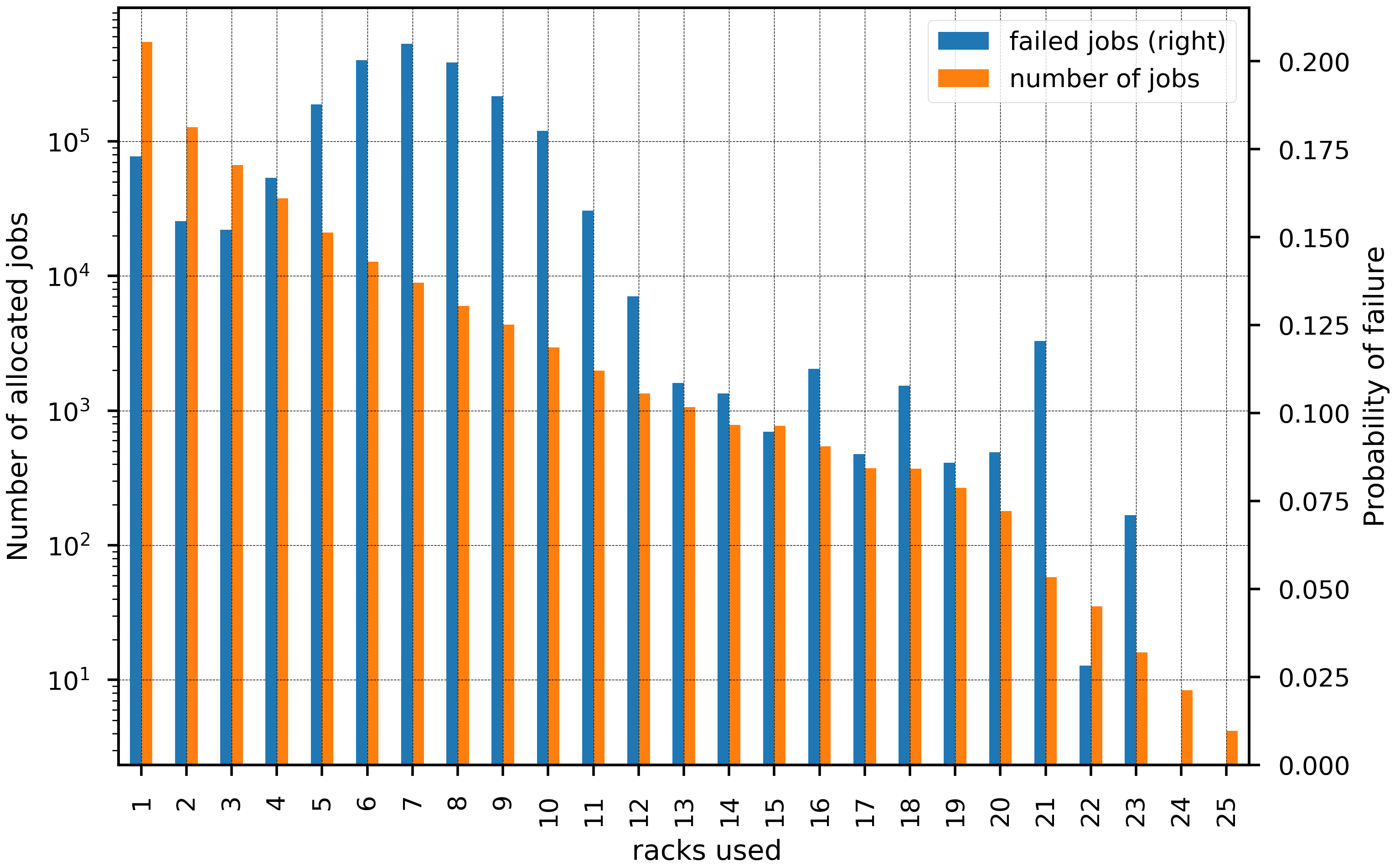}
\caption{Plot presenting number of racks used for allocations for all steps and failed steps. N=841k}
\label{jobs_vs_racks_v2}

\end{figure}

The mean number of racks used by multi-node steps is 1.92. This distribution is presented in Figure \ref{jobs_vs_racks_v2}. This figure also shows the probability of a failure according to the number of racks used for a step, and the maximum is at seven racks. For a number of racks over 13, which means using even more than 1000 computing nodes, occurrences of failures are rare. This phenomenon can be explained rather by a user's behavior than hardware dependencies. Most of HPC jobs are projected to be run on a specific number of nodes. This dependency is opposite to Big Data business software, where horizontal scaling on demand is one of the most important requirements in an application. So, the code for huge HPC jobs seems to be better tested and reliable for a fixed number of nodes.

\begin{figure}[!h]
\centering
\includegraphics[width=0.7\textwidth]{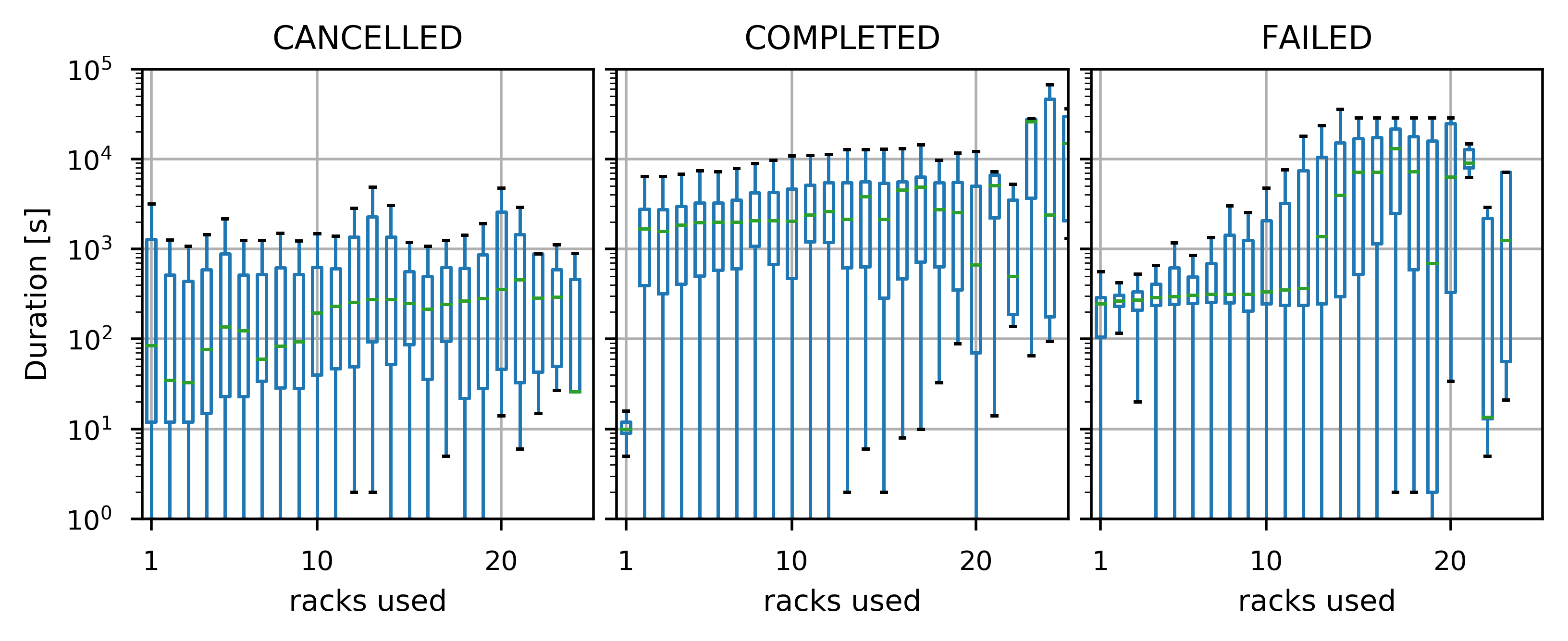}
\caption{Plot presenting duration depending on number of racks used for a step}
\label{racks_vs_duration}
\vspace{-30pt}
\end{figure}

\begin{figure}[!h]
\centering
\includegraphics[width=0.7\textwidth]{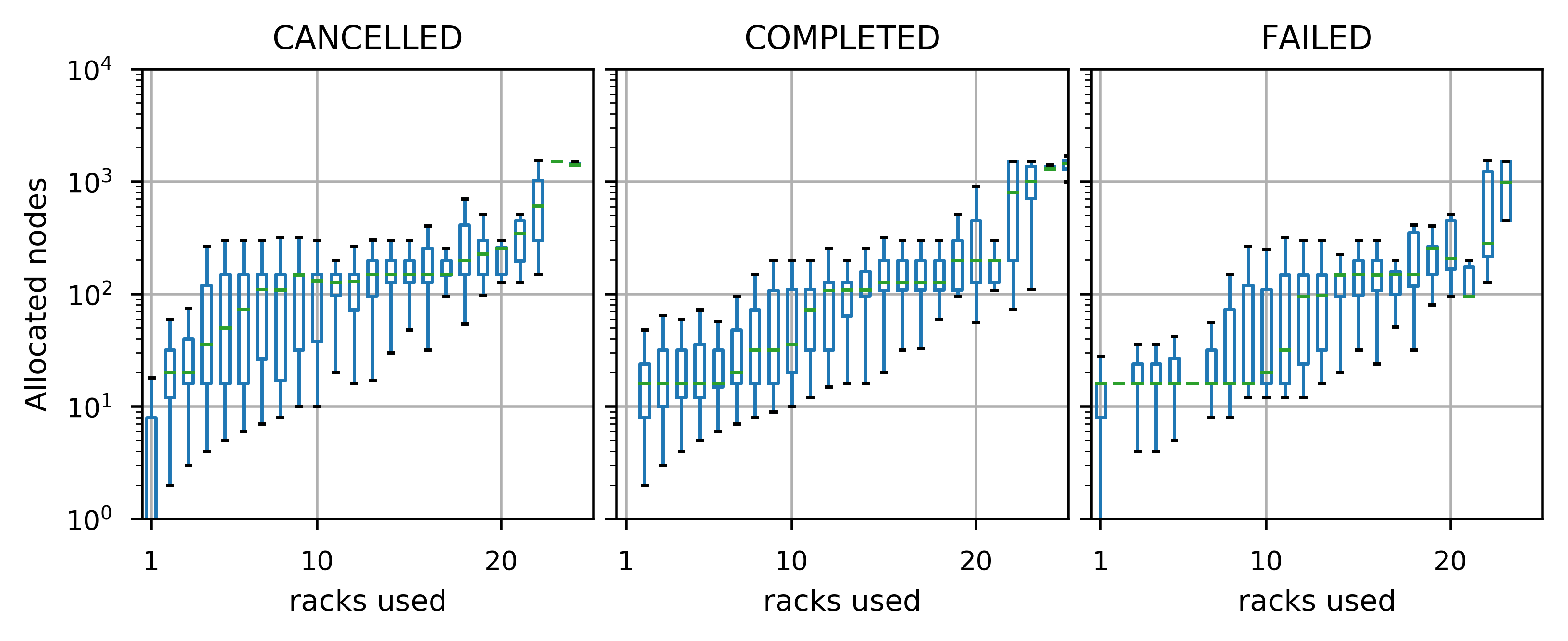}
\caption{Plot presenting number of allocated nodes depending on number of racks used for a step}
\label{racks_vs_number_nodes}
\vspace{-30pt}
\end{figure}

In Figure \ref{racks_vs_duration}, we analyze duration over a number of racks used by a step. Notably, failed steps are statistically shorter than completed, when approximately less than ten racks are used for a step. In this case, failures occur probably in the early phase of executed code. However, for a number of racks larger than 12, duration of failed steps significantly increases, while for completed ones it is kept on the same level. In Figure \ref{racks_vs_number_nodes}, distribution of the number of allocated nodes versus a number of racks can be seen. This relation is linear, although, in range of 10 and 20 racks used, the median number of allocated nodes does not increase. Cancelled steps with less than 100 nodes used are often placed in less than ten racks. It is opposite to failed or cancelled steps. These steps are more sparse, and for a few nodes allocated often use more racks.

\subsection{Node-power analysis}\label{subsection_node_power}
In Table \ref{table_powers_v1} we present average blade power and average last registered power for different job submissions states. 
In Table \ref{table_profile_power_state}, we present power statistics for steps longer than \(1000 s\), grouped by hardware profile. The table shows average values of power metrics in the last \(300 s\). It is seen, that for all hardware groups this value for failed steps is lower than for completed ones. Most probably, it is explained by the fact that once a software failure occurs some of the nodes go to an idle state. 
\begin{table}[h]
\centering
\caption{Power statistics depending on a submitted job state, for submissions longer than 120s}
\label{table_powers_v1}
\begin{adjustbox}{width=0.7\textwidth}
\begin{tabular}{|l|l|l|}
\hline
\textbf{Job finish state} & \textbf{Average blade power {[}W{]}} & \textbf{\begin{tabular}[c]{@{}l@{}}Average last registered\\ blade power {[}W{]}\end{tabular}} \\ \hline
\textbf{Completed}        & 265                                  & 228                                                                                            \\ \hline
\textbf{Failed}           & 242                                  & 227                                                                                            \\ \hline
\textbf{Cancelled}        & 240                                  & 203                                                                                            \\ \hline
\textbf{Node failed}      & 226                                  & 190                                                                                            \\ \hline
\end{tabular}
\end{adjustbox}
\end{table}

\begin{table}[h]
\caption{Avg power in last 300 s of a job, partitioned by a job and node, for jobs longer than 1000 s, then aggregated}
\centering
\label{table_profile_power_state}
\begin{adjustbox}{width=0.7\textwidth}

\begin{tabular}{|l|l|r|r|r|}
\hline
\textbf{Profile} & \textbf{State} & \textbf{\begin{tabular}[c]{@{}l@{}}avg(last\\ power\\ avg300)\end{tabular}} & \textbf{\begin{tabular}[c]{@{}l@{}}stddev(last\\ power\\ avg300)\end{tabular}} & \textbf{\begin{tabular}[c]{@{}l@{}}factor of \\ COMPLETED\end{tabular}} \\ \hline
\rowcolor[HTML]{EFEFEF} 
B720-compute\_36\_64 & CANCELLED & 196 & 76 & 1.11 \\ \hline
\rowcolor[HTML]{EFEFEF} 
B720-compute\_36\_64 & COMPLETED & 176 & 82 & 1.00 \\ \hline
\rowcolor[HTML]{EFEFEF} 
B720-compute\_36\_64 & FAILED & 172 & 71 & 0.97 \\ \hline
B720-compute\_36\_256 & CANCELLED & 209 & 75 & 1.00 \\ \hline
B720-compute\_36\_256 & COMPLETED & 210 & 82 & 1.00 \\ \hline
B720-compute\_36\_256 & FAILED & 186 & 76 & 0.89 \\ \hline
\rowcolor[HTML]{EFEFEF} 
B720-compute\_36\_128 & CANCELLED & 198 & 79 & 1.16 \\ \hline
\rowcolor[HTML]{EFEFEF} 
B720-compute\_36\_128 & COMPLETED & 170 & 82 & 1.00 \\ \hline
\rowcolor[HTML]{EFEFEF} 
B720-compute\_36\_128 & FAILED & 167 & 74 & 0.98 \\ \hline
B720-compute\_24\_64 & CANCELLED & 225 & 92 & 0.95 \\ \hline
B720-compute\_24\_64 & COMPLETED & 239 & 107 & 1.00 \\ \hline
B720-compute\_24\_64 & FAILED & 190 & 113 & 0.80 \\ \hline
\rowcolor[HTML]{EFEFEF} 
B720-compute\_24\_256 & CANCELLED & 269 & 134 & 0.97 \\ \hline
\rowcolor[HTML]{EFEFEF} 
B720-compute\_24\_256 & COMPLETED & 277 & 154 & 1.00 \\ \hline
\rowcolor[HTML]{EFEFEF} 
B720-compute\_24\_256 & FAILED & 242 & 154 & 0.87 \\ \hline
B720-compute\_24\_128 & CANCELLED & 248 & 116 & 0.93 \\ \hline
B720-compute\_24\_128 & COMPLETED & 266 & 141 & 1.00 \\ \hline
B720-compute\_24\_128 & FAILED & 227 & 142 & 0.85 \\ \hline
\end{tabular}
\end{adjustbox}
\end{table}

\textit{Additional analysis.} During this research we analyzed other issues, which are not presented in previous sections, but are valuable to notice. Firstly, we evaluated heat exchange between blades, to check if there is any correlation between the temperature of blades placed in the same chassis. Probably because of high-performance cooling infrastructure, no relationship is discovered. Another considered issue is the priority of a job submission in relation to its final state. No obvious correlation is observed, although an anomaly is detected in the distribution of priority level for the timeout state. Comparing to other states, a normalized frequency of submissions with high priority is significantly higher for timeouts.

\section*{Acknowledgment}
This research is supported by the BigStorage project (ref. 642963) founded by Marie Sk\l{}odowska-Curie ITN for Early Stage Researchers, and it is a part of a doctorate at UPC. 

\ifCLASSOPTIONcaptionsoff
 \newpage
\fi

\bibliographystyle{IEEEtran}

\end{document}